# A Novel VAPT Algorithm: Enhancing Web Application Security Trough OWASP Top 10 Optimization

Rui Ventura[1], Daniel José Franco[2] and Omar Khasro Akram[2]

[1] M.Sc., Faculty of Engineering, Institute Polytechnic of Beja, Portugal
[2] Asst. Prof., Faculty of Engineering, Institute Polytechnic of Beja, Portugal

*Abstract*

*This research study is built upon cybersecurity audits and investigates the optimization of an Open Web Application Security Project (OWASP) Top 10 algorithm for Web Applications (WA) security audits using Vulnerability Assessment and Penetration Testing (VAPT) processes. The study places particular emphasis on enhancing the VAPT process by optimizing the OWASP algorithm. To achieve this, the research utilizes desk documents to gain knowledge of WA cybersecurity audits and their associated tools. It also delves into archives to explore VAPT processes and identify techniques, methods, and tools for VAPT automation. Furthermore, the research proposes a prototype optimization that streamlines the two steps of VAPT using the OWASP Top 10 algorithm through an experimental procedure. The results are obtained within a virtual environment, which employs black box testing methods as the primary means of data acquisition and analysis. In this experimental setting, the OWASP algorithm demonstrates an impressive level of precision, achieving a precision rate exceeding 90%. It effectively covers all researched vulnerabilities, thus justifying its optimization. This research contributes significantly to the enhancement of the OWASP algorithm and benefits the offensive security community. It plays a crucial role in ensuring compliance processes for professionals and analysts in the security and software development fields.*

## 1. Introduction

Auditing cybersecurity for Web applications (WA) is crucial for organizations as a sizeable portion of their business is conducted online. Application attacks often occur due to the fast pace of software development, which can cause professionals to overlook security concerns. Ensuring security throughout the entire development and deployment lifecycle of WA can prevent high costs and efforts associated with implementing security measures in a later stage [1]. This investigation focuses on gaining an understanding of the current state of WA Cybersecurity Audits and VAPT, including the characterisation and concepts involved, as well as comprehending their methodologies and the tools utilized in these processes. Focused on the significance of implementing offensive security measures through VAPT for assessments of WA, along with modelling attacks by mapping the prevalent Web vulnerabilities [3]. This proposal includes the enhancement and optimization of an OWASP algorithm [2]. The focus of this study, and the enhance an OWASP algorithm, facilitates the formulation of a Main Research Question (Main-RQ): What can be optimise in the OWASP algorithm to facilitate the VAPT process in WA cybersecurity audits? And three research objectives (ROs):

- **RO1** - To characterize the audit of cybersecurity for Web and its automated tools, and suitable software tools used for automatic audits.
- **RO2** - To identify methods and techniques for automating Web cybersecurity audits with Pentest by using VAPT software tools; The tools employed in dynamically auditing the cybersecurity of WA.
- **RO3** - To optimize an OWASP algorithm that automates the VAPT procedure through software tools to address the modelling and mapping of prevalent WA vulnerabilities.

## 2. LITERATURE REVIEW

This section centres on the primary research areas of cybersecurity audits, VAPT, and OWASP algorithm optimisation that direct this literature review. It identifies the existing VAPT tools, methods, and processes, along with their connections. Additionally, it explores the potential for VAPT automation and the algorithms that permit automation, especially regarding OWASP. The aim of this section is to scrutinize the current data pertaining to the research areas and pinpoint the primary issues.

### 2.1. Cybersecurity Audits

#### 2.1.1. Cybersecurity Audits Concept

A cybersecurity audit offers an impartial outline of the techniques that organizations employ to examine and assess the overall technological and digital security stance of their assets [4]. The principal aim of the audit is to recognize, evaluate, and quantify any prospective risks. The objective is to detect and reveal any vulnerabilities that may lead to a data breach [5]. The term cybersecurity audit refers to an independent assessment and review of a system's records and activities. Its purpose is to evaluate the effectiveness of controls, ensure compliance with established security policies and procedures. According to [6], a cybersecurity audit has a comprehensive definition that underscores its importance in enhancing the security of modern information systems. According to earlier citations [4], [5], and [6], a cybersecurity audit aims to evaluate the risk level of a system or application. In brief, an audit assesses policies and controls, analyses, and measures information asset exposure, reports non-conformities, and proposes improvements systematically and continuously.

#### 2.1.2. Characteristics of Web Application Audit

According to [7], a security audit of WA, is a complex process made up of two fundamental areas: the type of audits to be carried out and the methodologies used to carry out the entire process of analysing vulnerabilities or non-conformities (see, Fig. 1). For [1] the security functions are related to confidentiality, integrity, availability, authentication, authorization, and non-repudiation.

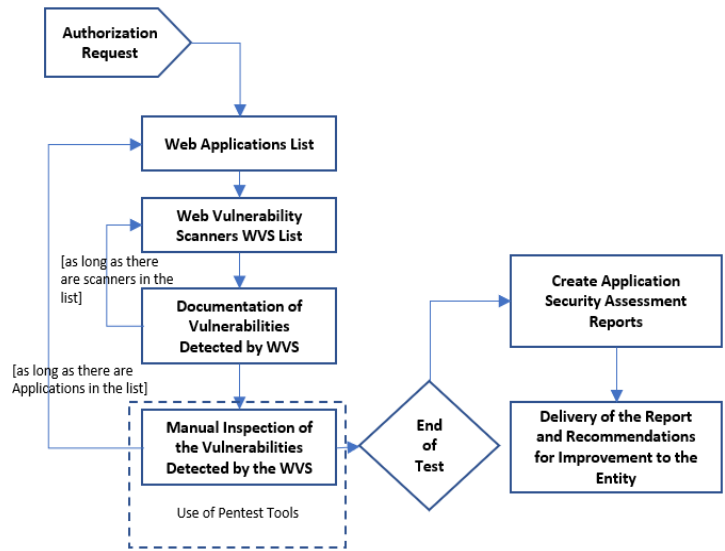

Figure 1. Web Application Auditing Methodology, adapted from [7].

From the perspective of offensive security at the service of application compliance, [8] argues that organizations should adopt Pentest as a way of simulating a real attack environment in a controlled manner using standards such as OWASP top 10. The Web Security Testing Guide (WSTG) is an internationally recognized guide from OWASP that provides characterisation and guidance on the various methodologies and techniques that can be used during an application audit using Pentest. The author [9] characterizes methods and techniques, considering the phase and course of the audit itself. According to the quotes from [1], [7], [8], both advocate the use of controlled intrusion tests (Pentest) to obtain application compliance, as for author [9], it supports and guides the analyst's own methods, techniques, and procedures while auditing WA and with an offensive security approach. In brief, a WA security audit is the process of testing, analysing, and reporting on the security level or posture of a WA, using standards, and obtaining an acceptable level of risk.

### 2.1.3. Automated Web Audit Tools

Web Application Vulnerability Scanners (WAVS) are automated tools that scan WA, typically from the user's perspective (client-side). These tools belong to the category of Dynamic Application Security Testing (DAST) tools. The majority of VA and PT tools are dynamic and automated in their processes. Automated software enables a more extensive analysis of both known and unknown vulnerabilities, as well as offering greater agility and consistency during the risk assessment process. [10–15]. According to NIST [16], the aim of a security or Web auditing tool is to establish a mandatory minimum level of functionality that both the buyer and supplier can use to verify the product. According to [17], analysis tools should be targeted and selected for a particular Web infrastructure and vulnerability. In short, cybersecurity auditing is the verification of information systems in each space and time, using tools, methodologies, standards, and objectives to assess and mitigate vulnerabilities. VAPT tools and standards such as OWASP, make it possible to audit and guarantee confidence in the security of WA.

## 2.2. Vulnerability Assessment and Penetration Testing (VAPT)

### 2.2.1. VAPT Processes

The VAPT process and life cycle can be seen as several sub-processes with various phases and techniques, which involve gathering, detecting, analysing information, and verifying the threat or weakness. In other words, it makes it possible to identify and assess vulnerabilities at a given phase and verifying their effectiveness or strength from the perspective of a Pentest or attacker in another. The vulnerability assessment (VA) process provides information about possible vulnerability while the penetration testing process includes exploiting the of the vulnerability to assume the level of risk [18]. For the author [22], VAPT is a process in which the Analyst or Pentester goes through several stages to explore a system or device and where objectives and some limitations are defined. According to [23], this type of approach aims to reduce FP and FN in the VAPT process. For these authors, the VAPT process is based on several phases using automated tools. Analysing and comparing the opinions of the authors [18], [22] and [23], both agree that VAPT is a continuous process and that its use enables the detection of vulnerabilities and identifying the risks, exploiting the threat. However, although they serve distinct functions and stages of an audit, when combined they provide real visibility of the risk and threat during an offensive security analysis or test. Based on the authors' review, it can be said that VAPT identifies vulnerabilities that can be exploited.

### 2.2.2. VAPT Automation

VA is a technique for detecting, categorizing, and prioritizing security vulnerabilities. Penetration Testing (PT) is a security technique that makes it possible to detect, verify and test security vulnerabilities [18]. However, the systematic combination of the two, through an automation or algorithm, can be seen as VAPT automation. According to [18], VAPT is a technique that combines the processes of detection, verification, and classification of vulnerabilities, combined with the knowledge of the analyst (Pentester). An automatic approach with VAPT involves scanning or analysing vulnerabilities dynamically, saving a lot of resources and contributing to greater consistency in the analysis. In a more targeted way and focused on a specific vulnerability, authors [19] created an automation that detects vulnerabilities based on the applications and services on the target system and specifically checks for the threat of Structured Query Language Injection (SQLI). Seeing the positive outlook for the automation of VAPT techniques, [20] designed an offensive security framework for smart homes using vulnerability analyses and penetration (VAPT) to efficiently protect data and devices throughout their lifecycle.

### 2.2.3. Pentest Automation

Automated Pentest is a vulnerability checking technique in which software is used to identify security flaws in information systems. It is seen as an alternative to manual Pentest, from the point of view of saving resources. For [21] the automation of penetration testing is to be able to combine various techniques and processes with a single objective. For the authors [21], is a simple and secure way of conducting all the test tasks. The authors [24], to automate the Pentest during an audit, developed a prototype system for executing automatic penetration tests, although the authors [24] describe the correct use of VAPT and its applicability, they do not mention any standard or methodology for the execution module. According to [25], the importance of Pentest WA demonstrated, as well as the importance of automating this process. VAPT automation is attractive, and you get the feeling that you only need a single tool to implement it, which is executed and generates the expected results. VAPT is an auditing process, manual or automated, with different objectives, methods, techniques, and tools.

## 2.3. Optimisation of OWASP Algorithms

### 2.3.1. OWASP top 10

The OWASP Top 10 framework is a widely accepted standard for software security, particularly for WA. It provides guidelines for developers on ensuring code security from writing to deployment. The framework identifies security risks as potential points of failure that developers should be aware of and mitigate. The Top 10 list classifies and lists these risks. In 2021, OWASP published 10 new WA security problems, including broken access control, cryptographic failures, injection, insecure design, security misconfigurations, vulnerable components, identification and authentication failures, security logging and monitoring failures, and server-side request forgery, aiming to educate programmers and analysts [26], [27].

### 2.3.2. Known Algorithms for the Pentest Process with OWASP

Research and literature review found few studies discussing the OWASP top 10 for automatic vulnerability analysis and exploitation. Pentest automation with OWASP top 10 is a software process that automates vulnerability verification through Pentest, improving the process based on a standard. The OWASP algorithm classifies attacks based on severity, frequency of defects, and impact [27]. To automate the most common steps in the Pentest process, [28] proposed a framework, these authors used a decision tree to select the best exploits. However, the authors [28] limit their research to the discovery of services and ports with two tools. To drive efficiency and practical correlation between OpenVAS and Metasploit, the authors [2] developed a system to automate the Pentest process from the result of a VA. These authors with the results obtained from Openvas, prioritize, and classify them with OWASP top 10, and test them automatically with Metasploit. The approach taken in the studies [2] and [28] is somewhat close to the intended objectives, particularly about modelling the attack with OWASP, but both show some limitations regarding the tools chosen for detecting and exploiting vulnerabilities in WA.

### 2.3.3. Algorithm Optimization OWASP top 10

Most VAPT related research focuses on identifying vulnerabilities at the network infrastructure level and relies mainly on technical procedures carried out manually by a Pentester or Security Analyst [29]. At this point, the focus is on work dedicated to automating the Pentest process and similar algorithms: Recently, there are some works [2], [28], [29] focused on the automation process, but both always use only two tools to automate the process. As for the authors [2], seen as an added value to this objective, they approach the exploration and prioritization of vulnerabilities based on the OWASP top 10 list. To evaluate 5 WAVS, [3] adopted a Black Box Testing method to analyse 7 WA. The analysis and evaluation focused on different metrics and a baseline of 8 vulnerabilities, mapped using NIST and OWASP. In brief, an OWASP top 10 algorithm can classify risks according to the severity of vulnerabilities, the frequency of isolated security defects and their possible impact. The automated processes based on this Standard, combined with the automation of VAPT, demonstrate its effectiveness while auditing WA

## 2.4. Theoretical Hypothesis

Theoretical hypothesis, and for this research a ladder diagram [30] is formulated in this process is a method that allows to assist and combine the ideas and summaries obtained through the main constructs and literature review. Based on the reviewed literature surrounding cybersecurity audits, tools, and automation for optimizing the VAPT process with the implementation of an OWASP algorithm, and the advancement of the study towards a proof of concept, these authors [29], [2], [28] closely align with the intended objective. However, the primary focus of this research study is prototype [2]. Therefore, it determined that for this research's experimental process methodology and evaluation, the benchmark [3] should be the preferred option. In brief, this section offers general concepts about the study, including its focus, primary research objectives, research methodology, and limitations. The assessed algorithms, which utilize OWASP classification and VAPT techniques, prioritize analysis of specific vulnerabilities or network infrastructure. This is attributed to the limited capability of the detection tools used in detecting or covering the majority of OWASP Top 10 vulnerabilities. Many investigations identified as lacking methodology. This can be attributed to the distinct methodologies and approaches adopted by individual analysts or researchers, with the aim of aligning their techniques with the practical objectives.

## 3. METHODOLOGY

This point highlights the methodology and methods adopted, focusing on a quantitative approach through an experimental method (see, Fig.2). As a result, this section provides information on how the study should be conducted to fulfil its initial objectives [30]. According to the nature of this investigation, a quantitative methodology, based on experimental methods, will be adopted a quantitative analysis to obtain detailed results and comparisons [30-32]. Quantitative research is appropriate for this study, as it allows us to the effectiveness and precision rate of the optimisation proposed.

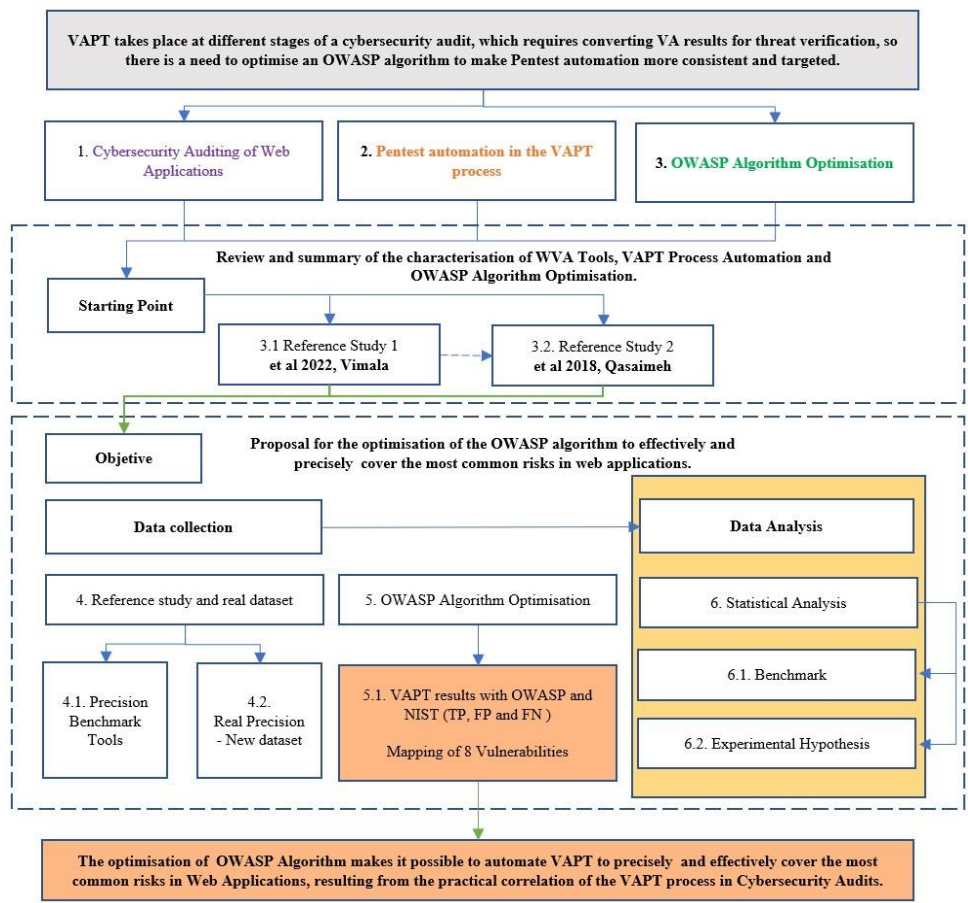

Figure 2. Research Framework and Methodology.

## 3.1. Experimental Design

### 3.1.1. Tools (WAVS)

For this experiment, 5 WAVS are used, together with the proposed prototype (see, Fig.3).

| Tools | Vendor | Version | Licence | Platform |
|---|---|---|---|---|
| Acunetix | Acunetix | 13.0.200217097 | Comercial | Windows, Linux |
| Burp Suite | Port Swinger | Pro v2022.8.2 | Comercial | Windows, Linux |
| Netsparker | Inviti | Pro 5.8.1.28119 | Comercial | Windows, Linux |
| Nessus | Tenable | Pro 10.5.1 | Comercial | Windows, Linux, Mac |
| ZAP | OWASP | 2.12 | Open-Source | Windows, Linux, Mac |
| WebVAPT | Author | 1.0 | Open-Source | Windows, Linux |

Figure 3. List of investigated WAVS and Proposed prototype.

### 3.1.2. Web Applications (WA)

To increase the granularity and diversity of the vulnerabilities to be detected, as well as having different WA technologies, the evaluation uses the capabilities of each scanner with seven vulnerable WA [3]. According to [3] these WA are intended to facilitate Web developers, security auditors, and Pentesters to hone their knowledge and testing expertise without any concerns [3].

### 3.1.3. Datasets

The acquisition of the two datasets, treated as the primary source of this research study, are obtained according to [3], to measure and evaluate the WAVS and the optimisation of the proposed prototype.

- **Dataset 1**: First data is obtained by collecting and analysing the WA, using the WAVS and the proposed prototype with a Black Box Testing approach.
- **Dataset 2**: Second data is obtained from a manual analysis and verification, considering the results obtained in the first set of data, validating the results of the first dataset, centred on the TP, FN, and FP rates.

### 3.1.4. Research Variables

To assess the optimisation of the algorithm proposed in this study, a complete randomization of the experiment design block will be used for which the method: 2 Factors with 6 Treatments. By choosing this randomization and research method, according to [30-32], it is possible to: Describe the statistical independence (analysis requirement); Randomly assign subjects, objects, and the order in which the tests are carried out; check the effect of other factors; and select objects that are representative of the population of interest.

### 3.1.5. Virtual Laboratory

The proposed algorithm, optimised using python programming language [34], as well as the rest of the WAVS, are executed and operated on a virtual platform (Virtual Machine) in Linux, Debian 12.2.0-14 with 8 CPU, 16GB RAM, 500GB SSD, Linux distribution version 6.1.0-kali7-amd64 [35]. Except for NetSparker, which used a virtual machine with the Windows 10 Pro operating system [36], with 4 CPU, 10GB RAM, 60GB SSD. Although the authors [3] do not mention or make any reference to the environment for collecting the two datasets, the results will

20 Computer Science & Information Technology (CS & IT)20                    Computer Science & Information Technology (CS & IT)

not be influenced since the data is generated by the same tools. For standard results, and in accordance with reference [3], the scanners are parameterised with default scanning settings. The base scanning profile is utilised without any customisations or adjustments.

### 3.2. Experimental Hypothesis

The proposed OWASP algorithm enables the automation and parameterization of VAPT tool processes and correlation to achieve greater consistency and precision in WA audits. To evaluate the precision of the algorithm during a WA audit, it will model 8 vulnerability types mapped with NIST and OWASP, aiming for a precision and detection rate of at least 90%. The rate of FP detected provides metrics to evaluate the precision [3] of the algorithm. The algorithm should make it possible to optimise and correlate VAPT tools, selecting a set of specific Vulnerability Assessment (VA) tools that will be parameterized and aimed to detecting the investigated vulnerabilities. In turn, Pentest (PT) will use exploitation tools that are targeted and orientated according to mapped vulnerabilities.

### 3.3. Research Metrics

The following metrics were identified to analyse and evaluate the WAVS tools and the proposed algorithm:

#### 3.3.1. True Positives Rate

The True Positives (see, Fig.4) are the proportion of vulnerabilities detected correctly by the WAVS and the proposed prototype.

$$Detection\ Rate = \frac{TP}{TP + FN}$$

Figure 4. True Positive (TP) Rate formula, [31].

#### 3.3.2. False Positives Rate

The False Positives (see, Fig.5), are the proportion of vulnerabilities detected incorrectly by the WAVS and the proposed prototype.

$$False\ Positive\ Rate = \frac{FP}{FP + TN}$$

Figure 5. False Positive (FP) Rate formula, [31].

#### 3.3.3. False Negatives Rate

False Negatives (see, Fig.6), are relate to vulnerabilities that were not detected by the WAVS and the proposed prototype, but which do exist.

$$False\ Negative\ Rate = \frac{FN}{FN + TP}$$

Figure 6. False Negative (FN) Rate formula, [31].

### 3.3.4. Precision Rate

The Precision Rate, (see, Fig.7) is a measure of the precision of a model or algorithm in predicting and classifying tasks. It is defined as the ratio of true positive predictions to the total number of positive predictions [33].

$$Precision\ Rate = \frac{TP}{TP+FP}$$

Figure 7. Precision Rate formula [33].

### 3.3.5. Efficacy Rate

This metric is focused on the detection rate, also known as the TP rate. This metric, in the context of evaluating the effectiveness of the WAVS tools and the proposed algorithm, focuses on the vulnerabilities not detected by the scanners (FN), [31].

## 3.4. Data Collection and Analysis

The collection and analysis of this experimental study is based on a Black Box Testing approach. The main benefit of this approach is to provide a scenario like real and more common attacks [3].

### 3.4.1. Data Collection

As a primary source of research and according, to [3] the methodology used to obtain the first dataset uses the selected scanners to analyse each of the seven WA to identify or detect possible vulnerabilities. For each WA, a report is generated by the respective scanner which lists the vulnerabilities detected. Among the analysis and evaluation criteria used in this study, the vulnerabilities to be investigated were extracted by mapping the NIST and OWASP (see, Fig.8). The purpose of the mapping is to identify overlapping vulnerabilities between the two standards. It also found that some vulnerabilities are resource-dependent, and others may require source code analysis to detect [3]. These types of vulnerabilities are eliminated since they cannot be verified from a Blackbox Testing perspective.

| NIST | | | OWASP | | Mapping results | | |
|---|---|---|---|---|---|---|---|
| V | M | D | V | D | V | D | Investigated Vulnerabilities |
| N1 | A7 | ✓ | A1 | ✓ | V1 | ✓ | Cross Site Scripting (XSS) |
| N2 | A1 | ✓ | A2 | ✓ | V2 | ✓ | Injection |
| N3 | A1 | ✓ | A3 | ✓ | V3 | ✓ | Broken Authentication |
| N4 | A1 | ✓ | A4 | ✗ | V4 | ✓ | Security Misconfiguration |
| N5 | - | ✓ | A5 | ✓ | V5 | ✓ | Sensitive Data Exposure |
| N6 | - | ✓ | A6 | ✓ | V6 | ✓ | Malicious File Inclusion |
| N7 | A9 | ✗ | A7 | ✓ | V7 | ✓ | Cross Site Request Forgery (CSRF) |
| N8 | A7 | ✓ | A8 | ✓ | V8 | ✓ | Insecure Communication |
| N9 | A3 | ✓ | A9 | ✗ | | | |
| N10 | A6 | ✓ | A10 | ✗ | | | |
| N11 | A2 | ✓ | | | | | |
| N12 | A6 | ✓ | | | | | |
| N13 | A2 | ✓ | | | | | |
| N14 | A5 | ✓ | | | | | |

*(V for Vulnerability, M for Mapping to OWASP, and D for Detectability by a scanner)*

Figure. 8. Investigated Vulnerabilities Mapping [3].

**3.4.2. Data Analysis**

This stage shows how the results were obtained, and the metrics considered. The total number of vulnerabilities detected by each scanner, namely the mapping of the 8 investigated vulnerabilities and the number of FP, FN and TP obtained by each scanner. The first research dataset is then subjected to manual analysis to verify FN and FP vulnerabilities. The FN are obtained to, identifying vulnerabilities not detected by scanners. FN, in this context, being of relevant importance for calculating and evaluating the effectiveness. In this research, the main sources of data are obtained through different approaches. It is possible to understand the steps and how the primary and secondary sources of this research study are obtained, their purpose and how they are dealt with. According to [30], this type of tactic guarantees a backdrop for the following stages of the research and provides the basis for framing and objectively optimizing the proposed research algorithm. It is possible to comprehend the procedures involved in obtaining primary and secondary sources for this investigation, as well as their respective purposes and how they are managed. To validate and compare the findings gathered from the second dataset (primary source) with those on [3], a statistical analysis will be conducted. In this investigation, inferential statistics will be utilized to derive a statement about a population based on a sample [31].

**3.4.3. Data Validation**

Data validation is achieved with the experiment, supported by the methodology used in [3], using the dataset 2 and obtaining the FP and TP rate results from the first dynamic analysis carried out by the WAVS and the prototype. In turn, a statistical comparison is made of the results obtained from the experiment with data and results from [3]. Due to the nature of this research and its objectives, it is necessary to analyse more than one source of data to obtain reliable assurance that the forementioned objectives are achieved. Furthermore, due to the nature of this research, where the aim is to prove an experimental hypothesis, according with author [31], inferential statistics help to answer the Main-RQ of this study by statistically validating the results.

## 4. RESULTS

The aim of this section, as directed in RO3, is to evaluate the results of the upgraded OWASP algorithm. This research displays the precision and effectiveness of the refined WebVAPT prototype. WebVAPT focuses on identifying 8 (eight) specific vulnerabilities: Cross-site scripting (XSS), Injection, Broken Authentication, Security Misconfiguration, Sensitive Data Exposure, Malicious File Inclusion, Cross Site Request Forgery (CSRF) and Insecure Communication. It also aims to create user-friendly, automated, and parameterized VAPT tools that can detect and exploit these vulnerabilities. The proposed algorithm aims to automate and parameterise VAPT tools, as shown in Figure 9, consisting of four modules. It allows a comprehensive analysis by technique, VA, or PT, and/or a combination of methods, VAPT, and individually by vulnerability or tool.

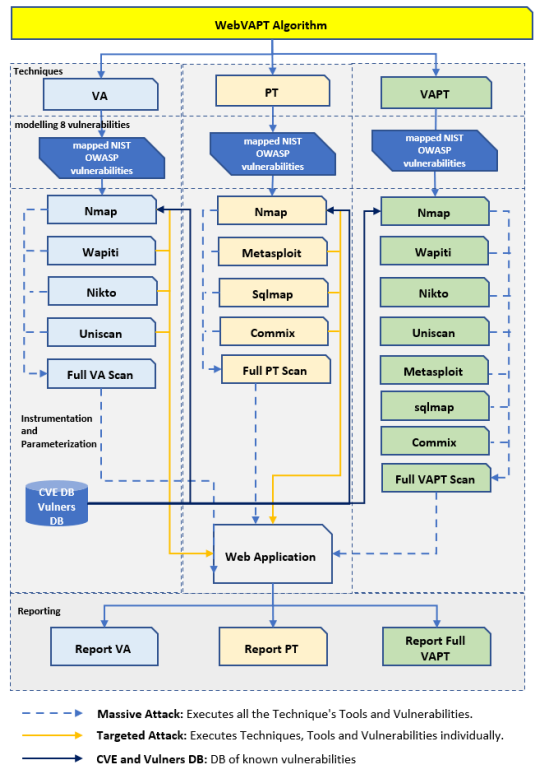

Figure. 9. Diagram of the proposed prototype.

## 4.1. Dataset 1

These results focus specifically on collecting, analysing, and detailing the data obtained in accordance with the research methodology, detailed in section 3. Figure 10 quantifies and illustrates the total number of vulnerabilities detected by the five WAVS, together with the WebVAPT prototype.

| Web Applications | Evaluated Tools by Application | | | | | |
|---|---|---|---|---|---|---|
| | Acunetix | Burp Suite | Netsparker | Nessus | ZAP | WebVAPT |
| W1 | 33 | 41 | 100 | 22 | 11 | 82 |
| W3 | 45 | 33 | 227 | 20 | 15 | 137 |
| W4 | 136 | 49 | 136 | 64 | 19 | 90 |
| W5 | 51 | 41 | 40 | 62 | 20 | 62 |
| W6 | 7 | 10 | 15 | 10 | 5 | 37 |
| W7 | 14 | 16 | 86 | 17 | 15 | 24 |

Figure 10. Total number of vulnerabilities detected by each tool and by WA.

It should be noted that the WA W2 was not available while the W6 was only partially available, displaying an HTTP 403 forbidden error. Nonetheless, it was examined for auditing as it was responsive on other HTTP ports like tcp/8080 and tcp/443. Benchmark [3] should be consulted for further details. The proposed prototype, as depicted in Figure 11, enables the comprehensive identification and coverage of prevalent vulnerabilities. Specifically, it maps and models the investigated vulnerabilities, with implications for all vulnerabilities.

| Vulnerabilities | Acunetix | BurpSuite | Netsparker | Nessus | Zap | WebVAPT |
|---|---|---|---|---|---|---|
| V1 | 29 | 20 | 35 | 10 | 20 | 51 |
| V2 | 30 | 24 | 59 | 14 | 12 | 81 |
| V3 | 4 | 0 | 0 | 1 | 0 | 3 |
| V4 | 116 | 33 | 99 | 85 | 97 | 121 |
| V5 | 35 | 39 | 58 | 56 | 101 | 62 |
| V6 | 3 | 0 | 2 | 0 | 0 | 1 |
| V7 | 41 | 5 | 36 | 0 | 35 | 43 |
| V8 | 22 | 12 | 27 | 8 | 4 | 46 |
| Total | 280 | 133 | 316 | 174 | 269 | 408 |

Figure 11. Total number, of investigated vulnerabilities, detected by each scanner.

The results illustrated indicate (see, Fig. 12), at this phase, that considerable progress has been made in identifying most of the vulnerabilities examined, apart from vulnerability V6 (Malicious File Inclusion). This limitation has resulted from the identification of certain security measures, like Web application Firewall (WAF), during data collection. Regarding vulnerability V4 (Security Misconfiguration) and the lowest percentage result achieved in this research, it is the opinion of this investigation that the WAVS has improved its ability to interpret application and server responses.

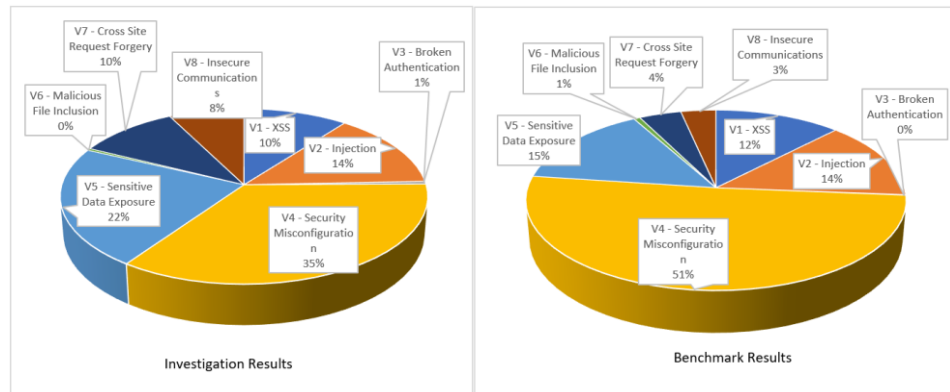

Figure 12. Investigated vulnerabilities detected by percentage. Comparison with benchmark.

### 4.2. Dataset 2

The second dataset is obtained by manually analysing the vulnerabilities detected in the first dynamic collection (see, Fig. 13). The manual analysis and re-analysis of the vulnerabilities previously detected dynamically makes it possible to validate whether the vulnerabilities were detected incorrectly or were not detected (FP and FN).

| Vulnerabilities | Acunetix | BurpSuite | Netsparker | Nessus | zap | WebVAPT |
|---|---|---|---|---|---|---|
| True Positives | 261 | 159 | 555 | 280 | 543 | 389 |
| False Positives | 21 | 23 | 49 | 38 | 64 | 33 |
| Total | 282 | 182 | 604 | 318 | 607 | 422 |
| Study precision | 93% | 87% | 92% | 88% | 89% | 92% |
| Benchmark Precision | 92% | 50% | 91% | 91% | 73% | |

Figure 13. Second Dataset. Obtained results and Precision Rate comparison to benchmark.

We focus on the validation of vulnerabilities reported as FP, for the purpose of calculating and evaluating the precision of the WAVS and the prototype, and on FN to assess effectiveness. Comparing with [3] the evaluation results obtained in this research study (see, Fig.14), we found that a percentage improvement in Acunetix and Netsparker of 1%. BurpSuite with 37% more, proving to be the one that improved its precision rate the most. OWASP Zap also improved its rate, by 16% and Nessus, decreased its precision rate, obtaining a rate of 88%, where [3] obtained 91%. As for proposed WebVAPT, showed a precision rate of 92% in detecting the investigated vulnerabilities.

|  | Acunetix | BurpSuite | Netsparker | Nessus | ZAP | WebVAPT |
|---|---|---|---|---|---|---|
| Obtained Precision Rate | 93% ▲ | 87% ▲ | 92% ▲ | 88% ▼ | 89% ▲ | 92% |
| Benchmark Precision Rate | 92% | 50% | 91% | 91% | 73% |  |

Figure 14. Comparison of the precision rate results of the investigation, with the benchmark.

Acunetix once again proved, here and in terms of effectiveness, the one that achieved the best results with a rate of 98%, and WebVAPT achieved a considerable effectiveness rate of 96% (see, Fig.15).

| Vulnerabilities | Acunetix | BurpSuite | Netsparker | Nessus | ZAP | WebVAPT |
|---|---|---|---|---|---|---|
| True Positives | 261 | 159 | 555 | 280 | 543 | 389 |
| False Negatives | 4 | 15 | 22 | 18 | 23 | 15 |
| Total | 265 | 174 | 577 | 298 | 566 | 404 |
| Efficiency rate | 98% | 91% | 96% | 94% | 96% | 96% |

Figure 15. Calculating Results and Analysing the Effectiveness Rate.

The results provided an updated investigation compared to a previous [3]. In addition, the study added value by measuring and analysing the effectiveness. The results showed an improvement in precision and a decrease in the FP Rate. Overall, the study has shown an enhancement in tool precision and has yielded positive results. In brief, the proposed WebVAPT algorithm can be deemed as a proficient approach, exhibiting significantly efficient and precise coverage of the investigated vulnerabilities.

## 5. CONCLUSION

The research emphasizes the significance of adhering to internationally recognized standards like NIST and OWASP in cybersecurity. It highlights the effectiveness of automating VAPT processes, enabling organizations to streamline security auditing efforts and enhance efficiency. Our optimized algorithm demonstrates high precision and detection capacity, which can reduce FP and FN in WA security audits, saving valuable time and resources. Furthermore, our research has implications for cybersecurity education and training. While the results align with our objectives, there is potential to improve vulnerability exploitation by using more efficient tools and refining associated payloads. Additionally, future research could explore the use of Artificial Intelligence or Machine Learning techniques to map vulnerabilities in the latest version of OWASP Top 10. These advancements would contribute to the continuous development of cybersecurity practices.

**ACKNOWLEDGEMENTS**

The research team would like to acknowledge the support of ONDAS Scholars for this project.

## AUTHORS

**Rui Ventura**, M.Sc. in Computer Security and Engineering Education at the Faculty of Engineering, at Polytechnic Institute of Beja, is currently working in cybersecurity field and law enforcement at Portuguese Government. Research interests include Cybersecurity Audits, Penetration Testing, Information Security Management and Network Security. His Certified as ISO 27001 lead and internal auditor, Bureau Veritas. Leading specialist trainer in Intrusion Protection Systems (IPS) by Hewlett-Packard. Certification of teaching competences through the Portuguese Employment and Vocational Training Institute (IEFP).

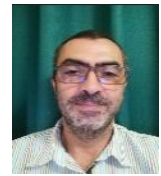

**Daniel José Franco**, Ph.D. introduced 'Reserchology' in the third decade of the 21st century. He holds positions as a senior researcher and assistant professor at the Institute Polytechnic of Beja in Portugal. He serves as the Chief Technology Officer (CTO) of the 'Apina's Foundation' in New York, USA. With a Ph.D. in computer networks from Universiti Putra Malaysia, he is an expert in his field. Committed to academia, he actively engages in teaching and supervising postgraduate students. His research interests encompass computer security, the Internet of Things, smart cities, and intelligent urban development. Additionally, he is recognized for his role in conducting workshops on research methodology. Please direct correspondence to: daniel.franco@ipbeja.pt.

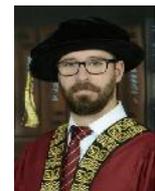

**Omar Khasro Akram**, Ph.D. pioneered the term 'Reserchology' in the third decade of the 21st century. He holds positions as a senior researcher and assistant professor at the Institute Polytechnic of Beja in Portugal. He serves as the Chief Operating Officer of the 'Apina's Foundation' in New York, USA. With a Ph.D. in urban planning and design from Universiti Putra Malaysia, he is an expert in his field. Driven by a passion for academia, he actively engages in teaching and supervising postgraduate students. His broad research interests encompass Reserchology, Conservation Management, Urban Heritage, Heritage-Smart Cities, Islamic Architecture, Theory and History of Architecture, and more. Additionally, he is recognized for his role in conducting workshops on research methodology. Please direct correspondence to: omar.akram@ipbeja.pt/ omar.khasro@yahoo.com

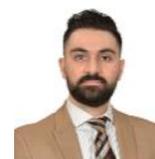